# Parallel STEPS: Large Scale Stochastic Spatial Reaction-Diffusion Simulation with High Performance Computers

**Weiliang Chen[1*], Erik De Schutter[1]**

[1]Computational Neuroscience Unit, Okinawa Institute of Science and Technology Graduate University

**\* Correspondence:**
Weiliang Chen
w.chen@oist.jp

## Abstract

Stochastic, spatial reaction-diffusion simulations have been widely used in systems biology and computational neuroscience. However, the increasing scale and complexity of simulated models and morphologies have exceeded the capacity of any serial implementation. This led to development of parallel solutions that benefit from the boost in performance of modern large-scale supercomputers. In this paper, we describe an MPI-based, parallel Operator-Splitting implementation for stochastic spatial reaction-diffusion simulations with irregular tetrahedral meshes. The performance of our implementation is first examined and analyzed with simulations of a simple model. We then demonstrate its usage in real-world research by simulating the reaction-diffusion components of a published calcium burst model in both Purkinje neuron sub-branch and full dendrite morphologies. Simulation results indicate that our implementation is capable of achieving super-linear speedup for balanced loading simulations with reasonable molecule density and mesh quality. In the best scenario a parallel simulation with 2000 processes achieves more than 3600 times of speedup relative to its serial SSA counterpart and more than 20 times of speedup relative to parallel simulation with 100 processes. While simulation performance is affected by unbalanced loading, a substantial speedup can still be observed without any special treatment.

# 1    Introduction

Active research in the fields of systems biology and computational neuroscience, such as the study on Purkinje cell calcium dynamics (Anwar et al., 2014), has significantly boosted the development of spatial stochastic reaction-diffusion simulators in recent years. These simulators split into two major categories according to the methods used, voxel-based and particle-based. Voxel-based simulators, such as STEPS (Hepburn et al., 2012), URDME (Drawert et al., 2012), MesoRD (Hattne et al., 2005), NeuroRD (Oliveira et al., 2010), divide the geometry into small voxels where different spatial variants of the Gillespie Stochastic Simulation Algorithm (Gillespie SSA) (Gillespie, 1976) are applied. Particle-based simulators, for example Smoldyn (Andrews and Bray, 2004) and MCell (Kerr et al., 2008), track the Brownian motion of individual molecules, and simulate molecule reactions according to their collisions. Although achieving great success, both voxel-based and particle-based approaches are known to be computational expensive. Particle-based simulators suffer from the requirement of tracking the position and movement of every individual molecule in the system. While tracking individual molecules is not required for voxel-based simulators, the exact solution of the Gillespie exact SSA  is known to be highly sequential and inefficient for large-scale simulation with massive numbers of SSA events (Dematté and Mazza, 2008).

There is a high need for more efficient stochastic spatial reaction-diffusion simulation of large scale models. Over the years many efforts have been devoted and achieved considerable successes, both in algorithm developments and software implementations, but the increasing simulation scale and complexity has significantly exceeded the speedup gained from these efforts.

Since the introduction of the original Gillespie SSA, the performances of voxel-based simulators have been substantially improved thanks to new algorithms and data structures. Giving $N$ as the number of kinetic events (reactions and diffusions) in the system, the computational complexity of a single SSA iteration has been reduced from O($N$) with the Direct method (Gillespie, 1976), to O(log2($N$)) with Gibson and Bruck's modification (Gibson and Bruck, 2000), to O(1) with the composition and rejection SSA (Slepoy et al., 2008). Approximate solutions for well-stirred systems such as the well-known tau-leaping method (Gillespie, 2001) can also be applied to spatial domain (Koh and Blackwell, 2011; Marquez-Lago and Burrage, 2007), which provide further speedups with controllable errors. It is clear, however, that the performance of a serial simulator is restricted by the clock speed of a single computing core, while multi-core CPU platforms have become main stream.

One possible way to bypass the clock speed limitation is parallelization, but the development of an efficient and scalable parallel solution has proven to be challenging. An optimistic Parallel Discrete Event Simulation (PDES) solution has been applied to the exact Gillespie SSA, achieving maximum 8 times of speedup with a 12 core cluster (Dematté and Mazza, 2008). This approach has been further investigated and tested with different synchronization algorithms available for PDES systems (Wang et al., 2009), such as Time Warp (TW), Breathing Time Bucket (BTB) and Breathing Time Warp (BTW). Their results indicate that while considerable speedup can be achieved, for example 5 times of speedup with 8 cores using the BTW method, the speedup decays rapidly once inter-node communication is involved, due to significant network latency. Another optimization attempt of the PDES solution with thread-based implementation has been reported lately (Lin et al., 2015), which achieved 9 times of speedup with 32 processing threads. All above studies show problematic scalability, due to the dramatic increase of rollbacks triggered by conflicting diffusion events between partitions, even with the support from well-developed PDES algorithms.

Parallelization of approximate SSA methods has also been investigated. D'Agostino and colleagues (D'Agostino et al., 2014) introduced a parallel spatial tau-leaping solution with both Message Passing Interface (MPI) based and Graphics Processing Unit (GPU) based implementations, achieving 20 times of speedup with a 32 node cluster, and about 50 times of speedup on a 192 core GTX-Titan. Two variants of Operator-Splitting approach, originating from the serial Gillespie Multi-Particle (GMP) method (Rodriguez et al., 2006), have been independently introduced by Roberts (Roberts et al., 2013) and Vigelius (Vigelius et al., 2011) to their GPU implementations, both achieve above hundred times of speedup comparing to serial SSA implementation. It is worth noticing that all above parallel solutions divide the simulated geometries into sub-volumes using a cubic mesh grid, which may not be suitable to accurately represent realistic morphologies.

Several studies of parallel particle-based implementations have been reported. Balls and colleagues (Balls et al., 2004) demonstrate their early attempt of parallel MCell implementation under KeLP infrastructure (Fink et al., 1998) with a 64 core cluster. Two GPU-based parallel implementations of Smoldyn have also been reported (Dematté, 2012; Gladkov et al., 2011), both show 100 ~ 200 times of speedup comparing to the CPU-based serial Smoldyn implementation.

Here we introduce an MPI-based parallel implementation of the STochastic Engine for Pathway Simulation (STEPS) (Hepburn et al., 2012). STEPS is a GNU licensed, stochastic spatial reaction-diffusion simulator implemented in C++ with a Python user interface. The main solver of current serial STEPS simulates reaction and diffusion events by applying a spatial extension of the composition and rejection SSA (Slepoy et al., 2008) to sub-volumes of unstructured tetrahedral meshes. Our parallel implementation aims to provide an efficient and scalable solution that can utilize state-of-art supercomputers to simulate large scale stochastic reaction-diffusion models with complex morphologies. In Section 2 we explain the main algorithm and details that are essential to our implementation. In Section 3, we then showcase two examples, from simple model to complex real-world research model, and analyze the performance of our implementation with their results. Finally, we will discuss possible future developments of parallel STEPS in Section 4.

## 2    Methods

We choose the MPI protocol on CPU clusters as the development environment of our parallel implementation, since it is currently the most well-supported parallel environment in academic research. Modern clusters allow us to explore the scalability of our implementation with massive number of computing nodes, and provide insightful information for further optimization towards super large scale simulations. The MPI-based implementation also serves as the foundation of future implementations with other parallel protocols and hardware such as GPU and Intel Xeon Phi clusters.

Previous attempts (Dematté and Mazza, 2008; Lin et al., 2015; Wang et al., 2009) of the exact Gillespie SSA parallelization have shown that system rollbacks triggered by straggler cross-process diffusion events can critically negate any performance gained from parallelization. The issue is further intensified for MPI-based implementation due to the significance of network latency. To take full advantage of parallelization, it is important to relax the exact time dependency of diffusion events and take an approximate, time-window approach that minimizes data communication and eliminates system rollbacks. Inspired by the GMP method, we developed a tetrahedral based Operator-Splitting algorithm as the fundamental algorithm of our parallel implementation. The serial implementation of this algorithm and its accuracy have been discussed previously (Hepburn et al., 2016). Here we discuss implementation details of the parallel version.

2.1 Initialization of a Parallel STEPS Simulation

To initialize a parallel STEPS simulation, the user is required to provide the biochemical model and geometry to the parallel solver. For the purpose of user convenience, our parallel implementation accepts the same biochemical model and geometry data used in the serial SSA solver as inputs. In addition, partitioning information of the mesh is required so that tetrahedrons can be distributed and simulated. The partitioning information is a simple list which can be generated automatically using the grid based partitioning solution provided in the STEPS utility module, or more sophisticated, third-party partitioning applications such as Metis (Coupez et al., 2000). The STEPS utility module currently provides necessary support functions for format conversions between STEPS and Metis files.

Assuming that a list of tetrahedrons is hosted by an MPI process $p$, $\{tet \mid tet$ is hosted by $p\}$, parallel STEPS firstly creates a standard Gillespie SSA system for all reactions in each hosted tetrahedron. This includes the population state of all species and reaction propensities. For each of the reaction $R_{tet,p}$, it also creates an update dependency list deps($R_{tet,p}$), that is, the list of reactions and diffusions that require an update if $R_{tet,p}$ is chosen and applied by the SSA. As a reaction only affects molecule states and propensities of reactions and diffusions within its own tetrahedron, the above information can be stored locally in $p$. The localized storage of SSA and dependency information significantly reduces the memory consumption for each process comparing to the serial SSA implementation, which is crucial to the simulator's performance. We will further address its importance with simulation results in section 3.

The simulation also stores the list of hosted diffusions $\{D_{tet,p} \mid D_{tet,p}$ is in $tet$ hosted by $p\}$ and the dependency list deps($D_{tet,p}$) for each diffusion $D_{tet,p}$. In addition, if a tetrahedron $tet$ is a boundary tetrahedron of $p$, in other words, the species state of $tet$ is affected by diffusions in tetrahedrons hosted by other MPI processes rather than $p$, a species update dependency list for every diffusive species $S_{tet,p}$ in $tet$ is also created. The species update dependency list, deps($S_{tet,p}$), is defined as the list of reactions and diffusions which are hosted by $p$, and require update if the count of $S_{tet,p}$ is modified by cross-process diffusion. The species dependency list allows each MPI process to update hosted reactions and diffusions independently after receiving molecule change information from other processes, thus reducing the need for cross-process communication.

Furthermore, a suitable diffusion time window is determined according to the biochemical model and geometry being simulated (Hepburn et al., 2016). Given $d_{S,tet}$ as the local diffusion rate for diffusive species $S$ in tetrahedron $tet$, each process $p$ computes a local minimal time window $\tau_p = min_{S,tet}(1.0/d_{S,tet})$ , over all diffusive species in every hosted tetrahedron. A collective communication is then performed to search for the global minimum, $\tau = min_p(\tau_p)$, which is set as the diffusion time window for every process in the simulation. It is worth mentioning that $\tau$ is completely determined by the biochemical model and geometry and stays constant regardless the changes in molecule population. Therefore no continuous update of $\tau$ is required during the simulation.

The final step is to initialize the molecule population state of the simulation, which can be done using various API functions provided in parallel STEPS. Once this is completed, the simulation is ready to enter the runtime main loop described below.

2.2 Runtime Main Loop

The runtime main loop for each MPI process is shown in Algorithm 1. When a process is asked to execute the simulation from time $t$ to $t_{end}$, a remote change buffer for cross-process data communication is created for each of the neighboring processes of $p$. Details of the buffer will be discussed later.

The entire runtime $[t, t_{end}]$ is divided into iterations of constant time window $\tau$, whose value is computed during initialization. At the start of every time window, each process first executes the Reaction SSA operator for the period of $\tau$. In addition to the standard exact SSA routines, the process also keeps tracking the update time and occupancy for each reactant and product species of reactions (Hepburn et al., 2016). The mean occupancies of species during the period of $\tau$ are used later to compute the number of molecules that diffuse to neighboring tetrahedrons in the Diffusion operator.

The parallel solver treats diffusion events differently based on the ownerships of tetrahedrons involved. If both the source and designation tetrahedrons of a diffusion event are in a single process, the diffusion is applied directly. If a diffusion is cross-process, that is, the source tetrahedron and designation tetrahedron are hosted by different processes, the change to the source tetrahedron is applied directly, while the change to the designation tetrahedron is registered to the corresponding remote change buffer. Once all diffusion events are applied or registered, the buffers are sent to designation neighboring processes where molecule changes are applied remotely.

The algorithm is designed for optimal operation in a parallel environment. Most of its operations can be performed independently without network communication. In fact, the only data communication required is the transfer of remote change buffers between neighboring processes. This leads to two important implications. First and foremost, the communication is strictly regional, meaning that each process only communicates to a small subset of processes that it shares geometry boundaries with, regardless of the overall scale of the simulation. Secondly, thanks to the non-blocking communication, each process can start the reaction SSA Operator for next iteration $t_1$, as soon as it has received remote change buffers for current iteration $t_0$ from all its neighboring processes and applied those changes (Figure 1). Therefore, data communication can be hidden behind computation, which helps to reduce the impact of network latency.

Since the remote change buffer is the only data being transferred across the network, it is important to limit its size so that communication time can be reduced. Furthermore, an efficient registering method is also required since all molecule changes applied to remotely hosted tetrahedrons need to be registered. Instead of recording every cross-process diffusion event, the remote change buffer records the accumulated change of a molecule species in a tetrahedron hosted remotely. Thus the size of the remote change buffer is upper-bounded by the number of remotely-hosted neighboring tetrahedrons and the number of diffusive species within those tetrahedrons. Each process also stores mirror images of remotely-hosted neighboring tetrahedrons. Each mirror image contains information of the neighboring tetrahedron's host process $p'$, its index on the hosted process $tet'$, indices of all diffusive species $S$ within, and a location marker $Loc_{tet',S}$ for each of the diffusive species. Correspondingly, the remote change buffer stores entries of changes one by one in a vector, where each entry consists of three elements, $tet'$, $S$, as well as the accumulated molecule change $m_{tet',S}$. When a new diffusion event causes molecule change of species $S$ in $tet'$ of $p'$, the host process of source tetrahedron first checks the entry data at $Loc_{tet',S}$ of the remote change buffer for $p'$. If the entry's tetrahedron index as well as species index matches the information of $tet'$ and $S$, the accumulated change of this entry is increased according to the diffusion event. As each buffer is reset after its content has been sent to corresponding process, a mismatch of entry information indicates such reset has taken place since previous registration of the same diffusion event, in which case a new entry ($tet'$, $S$, $m_{tet',S}$) is appended to the end of the buffer and the location of this entry is stored back to the location marker $Loc_{tet',S}$ for future reference. With a suitable data structure, both accessing entry data and appending new entry can achieve constant complexity, providing efficient solution for registering remote molecule changes.

## 3    Results

As the accuracy of the solution has been examined in a previous paper (Hepburn et al., 2016), here we mainly focus on the performance and scalability of our implementation. Simulations reported in this paper were run on OIST's university-shared high performance cluster, "Sango". Each computing node on Sango has two 12-core 2.5GHz Intel Xeon E5-2680v3 processors, sharing 128GiB of system memory. All nodes are interconnected using 56Gbit/s InfiniBand FDR. Due to the sharing policy only a limited number of cores could be used for our tests, and cluster conditions were different for each test and may scatter across the entire cluster. It is clear to us that cluster condition inevitably affects simulation performance. To accommodate this impact and provide important insights of how our implementation performs under real-life cluster restrictions, we repeated the tests multiple times, each starting at a different time and date with varying cluster condition, and report the averaged results.

Simulation performance were measured by both speedup and efficiency. Each simulation was run for a predefined period, and the wall-clock time was recorded. The averaged wall-clock time for a set of repeated simulations is hereby noted as $T_p$, where $p$ is the number of MPI processes used in each simulation. The speedup of a parallel simulation with $p$ processes relative to the one with $q$ processes is defined as $S_{p/q} = T_q / T_p$. Specifically, the speedup of parallel simulation with $p$ processes relative to its serial SSA counterpart is defined as $S_{p/SSA} = T_{SSA} / T_p$, where $T_{SSA}$ is the wall-clock time for the same simulation run by the serial SSA solver.

Correspondingly, we define the efficiency of a simulation with $p$ processes relative to the one with $q$ processes as $E_{p/q} = S_{p/q} \cdot \frac{q}{p}$. The efficiency measurement is commonly used to study the scalability of a parallel implementation. Specifically, the strong scalability of an implementation is measured by calculating the efficiency of parallel simulations of a fixed size problem with increasing number of computing processes, while weak scalability measures the efficiency of simulation whose overall problem size increases in direct proportion to the number of computing processes. Both scalabilities of our implementation were investigated here.

### 3.1 Reaction-Diffusion Simulation with Simple Model and Geometry

We first examine the simulation results of a fixed size reaction-diffusion problem. The simulated model was used to benchmark our serial spatial SSA solver in previous paper (Hepburn et al., 2012) and to test accuracy of our serial Operator-Splitting solution (Hepburn et al 2016), which consists of 10 diffusive species, each with differing diffusion coefficients and initial molecule counts, and 4 recursive reactions with various rate constants. The model was simulated in a $10 \times 10 \times 100 \mu m^3$ cuboid mesh with 3363 tetrahedrons. Tetrahedrons were linearly partitioned based on the y and z coordinates of their barycenter, where the numbers of partitions of each axis for a simulation with $p$ processes was arranged as [Parts$_x$ = 1, Parts$_y$ = 5, Parts$_z$ = $p/5$]. At the beginning of each simulation, species molecules were placed uniformly into the geometry, and the simulation was run for $t_{end}$ = 20 seconds, after which the wall-clock time was recorded. We started each series of simulations from $p$

= 5 and progressively increased the number of processes by 5 each time until $p$ = 300. Each series was repeated 30 times to produce an average result.

Speedup and efficiency are reported relative to the simulation result with 5 processes, in other words, $S_{p/5}$ and $E_{p/5}$. By increasing the number of processes, simulation performance of the fixed size problem improves dramatically. In fact, the simulation maintains super-linear speedup until $p \approx$ 250 (Figure 2A). While the efficiency decreases in general, it remains above 0.8 with $p$ = 300 (Figure 2B), where on average each process hosts approximately 10 tetrahedrons.

In addition to the overall wall-clock time, we also recorded the time cost of each algorithm segment to analyze the behavior of the implementation. The whole time cost for the simulation $T_{total}$ is divided into three portions. The computation time $T_{comp}$ includes the time cost for reaction SSA and the cost of diffusion operations within the process. The synchronous time $T_{sync}$ includes the time cost for receiving remote change buffers from neighboring processes, and the time cost for applying those changes. The time spent on waiting for the buffer's arrival, as well as the waiting time for all buffers to be sent after reaction SSA is recorded as the idle time, $T_{idle}$. In summary,

$$T_{total} = T_{comp} + T_{sync} + T_{idle}$$

A detailed look at the time cost distribution of a single series trial (Figure 2C) suggests that the majority of speedup is contributed by $T_{comp}$, which is consistently above the theoretical ideal (Figure 2D), thanks to the improvement of memory caching performance caused by distributed storage of SSA and update dependency information mentioned above. The result shows that $T_{sync}$ also decreases significantly as the number of processes increases, however, as the number of boundary tetrahedrons are limited in the simulations, $T_{sync}$ has the smallest proportion in overall time consumption (Figure 2C). Another important finding is that the change of $T_{idle}$ becomes insignificant when $p > 100$. Since $T_{comp}$ and $T_{sync}$ decrease as $p$ increases, $T_{idle}$ progressively becomes critical in determining the simulation performance. Other trials of simulations exhibited similar results but are not shown here.

To further study how molecule density affects simulation performance, we repeat the above test with two new settings, one reduces the initial count of each molecule species by 10 times, and the other increases molecule counts by 10 times (Figure 3A). We name these tests "Default", "0.1x" and "10x" respectively. The speedups relative to the serial SSA counterparts $S_{p/SSA}$ are also reported for comparison (Figure 3B). The 0.1x Simulations receive less performance boost from parallelization comparing to the default and 10x simulations. This is because in the 0.1x simulations $T_{comp}$ quickly decreases below $T_{idle}$, and the speedup becomes less significant as $T_{idle}$ is mostly consistent throughout the series (Figure 3C). In the 10x simulations $T_{comp}$ maintains its domination of the three timing segments, thus simulations achieve similar speedup ratio as the default ones (Figure 3D). This result also indicates that $S_{p/SSA}$ greatly depends on molecule density. In general, parallel simulation with high molecule density and high number of processes can achieve higher speedup relative to the serial SSA counterpart (Figure 3B).

Mesh coarseness also affects simulation performance greatly. Figure 4 shows the results of simulations with the same model and geometry dimensions, but different numbers of tetrahedrons within the meshes. Simulations with a finer mesh generally take longer to complete because there are more diffusion events, however, the time increment is not directly proportional to the increment of tetrahedron count. In fact, increasing the tetrahedron count from 13,009 to 113,096 only increases the wall-clock time for approximate 24% with $p$ = 300, while the simulation with 13,009 tetrahedrons takes 233% more time than the 3,363 tetrahedrons case (Figure 4B). Figure 4C shows that both 13,009 and 113,096 cases achieve dramatic relative speedups from parallelization, which are greatly above the theoretical ideal. This is because simulation with fine mesh and low process count has a high memory footprint per process and is unfriendly to memory caching. Massive parallelization greatly reduced the memory footprint for each process, thus improving the computational efficiency thanks to the caching effect.

Weak scalability can be studied by measuring the efficiency of a simulation whose size increases in direct proportion to the number of processes. We use the "Default" simulation with 300 processes as the baseline, and increase the problem size by duplicating the geometry along a specific axis, as well as increasing the number of initial molecules proportionally. Table 1 gives a summary of all simulation settings. As the problem size increases, the simulation efficiency progressively deteriorates (Figure 5). While around 95% of efficiency is maintained for doubling the problem size, tripling the problem size reduces the efficiency to about 80%. This is an expected outcome of the current implementation, because although the storage of reaction SSA and update dependency information are distributed, each process in the current implementation still keeps the complete information of the mesh geometry. Therefore the memory footprint per process increases as problem size increases, reducing the simulation efficiency. Optimizing memory footprint for super large scale problems is one of the main focuses in our next development iteration. This result also indicates that geometry partitioning plays an important role in determining simulation performance, as extending the mesh along the z axis gives better efficiency than along the y axis, even though they have similar numbers of tetrahedrons. This can be explained by the increase of boundary tetrahedrons in the latter cases. Since the number of boundary tetrahedrons determines the upper-bound of the size of remote change buffer and consequently the time for communication, reducing the number of boundary tetrahedrons is a general recommendation for geometry partitioning in parallel STEPS simulations.

3.2 Large Scale Reaction-Diffusion Simulation with Real-World Model and Geometry

Simulations from real-world research often consist of reaction-diffusion models and geometries that are notably more complex than the ones studied above. As a preliminary example, we extracted the reaction-diffusion components of a previously published spatial stochastic calcium burst model (Anwar et al., 2013) as our test model to investigate how our implementation performs with large scale real-world simulations. The extracted model consists of 15 molecule species, 8 of which are diffusive, as well as 22 reactions. Initial molecule concentrations, reaction rate constants and diffusion coefficients were kept the same as in the published model.

The Purkinje cell sub-branch morphology, published along with the model, was also used to generate a tetrahedral mesh that is suitable for parallel simulation. The newly generated mesh has 111,664 tetrahedrons, and was partitioned using Metis and STEPS supporting utilities. As discussed before, reducing boundary tetrahedrons is the general partitioning strategy for parallel STEPS simulations. This is particularly important for simulations with a tree-like morphology since grid based partitioning approach used for the simple cuboid above cannot capture and utilize spatial features of such morphology. The sub-branch mesh for our simulation is partitioned based on the connectivity of tetrahedrons. Specifically, a connectivity graph of all tetrahedrons in the mesh was presented to Metis as input. Metis then produced a partitioning profile which met the following criteria. First of all, the number of tetrahedrons in each partition was similar. Secondly, tetrahedrons in the same partition were all connected. Finally, the average degree of connections is minimum. Figure 6 shows the mesh itself as well as two partitioning profiles generated for $p = 50$ and $p = 1000$. As a preliminary test, this partitioning procedure does not account for any size differences of tetrahedrons and the influence from the biochemical model and molecule concentrations, although their impacts can be significant in practice. At present, some of these factors can be abstracted as weights between elements in Metis, however, substantial manual scripting is required and the solution is project-dependent.

To mimic the calcium concentration changes caused by voltage-gated calcium influx simulated in the published results (Anwar et al., 2013), we also extracted the region-dependent calcium influx profile from the result data, which can be applied to the parallel simulation periodically. Depending on whether this profile is applied, the parallel simulation behaved differently. Without calcium influx, the majority of simulation time was spent on diffusion events of mobile buffer molecules. As these buffer molecules were homogeneously distributed within the mesh, the loading of each process was relatively balanced throughout the simulation. When the calcium influx was applied and constantly updated during the simulation, it triggered calcium burst activities that rapidly altered the gradient of calcium concentration, consequently unbalancing the process loadings. It also activated calcium-dependent pathways in the model and increased the simulation time for reaction SSA operations.

Two series were simulated, one without calcium influx and data recording, and the other one with the influx enabled and data recorded periodically. Each series of simulations started from $p = 50$, and finished at $p = 1000$, with an increment of 50 processes each time. Both series of simulations were run for 30ms, and repeated 20 times to acquire the average wall-clock times. For the simulations with calcium influx, the influx rate of each branch segment was adjusted according to the profile every 1ms, and the calcium concentration of each branch was recorded to a file every 0.02ms, as in the original simulation. Figure 7A shows the recorded calcium activity of each branch segment over a single simulation trial period, which exhibits great spatial and temporal divergences. As a consequence of the calcium influx changes, process loading of the series was mostly unbalanced so that simulation speedup and efficiency were significantly affected. However, a substantial improvement can still be observed (Figure 7B and C). To further improve the performance of simulations with strong concentration gradients, a sophisticated and efficient dynamic load balancing algorithm is required, which will be briefly discussed later.

Finally, to test the capability of our implementation for full cell stochastic spatial simulation in the future, we generated a mesh of full Purkinje dendrite tree from a publically available surface reconstruction (3DModelDB (McDougal and Shepherd, 2015), ID: 156481) and applied the above model on it. To our best knowledge this is the first parallel simulation of mesoscopic level, stochastic, spatial reaction-diffusion system with full cell dendritic tree morphology. The mesh consisted of 1,044,155 tetrahedrons. As the branch diameters of the original reconstruction have been modified for other purposes, the mesh is not suitable for actual scientific study but only to evaluate computational performance. Because of this reason, and the fact that no calcium influx profile can be acquired for this reconstruction, we only simulated the one without calcium influx. The simulation series started from $p = 100$, and progressively increased to $p = 2000$ by an increment of 100 processes each time. The maximum number of processes ($p = 2000$) was determined by the fair-sharing policy of the cluster. Figure 8A gives an overview of the full cell morphology as well as a zoom-in look of the mesh. Both speedup and efficiency relative to simulation with $p = 100$ (Figure 8, B and C) show super-linear scalability and reach a peak with $p = 2000$. This result suggests that simulation performance may be further improved with a higher number of processes.

It is also worth mentioning that all parallel simulations above perform drastically better than their serial SSA counterparts. The speedups relative to the serial SSA simulations are shown in Figure 9. Even in the most realistic case with dynamically updated calcium influx as well as data recording, without any special load balancing treatment, the parallel simulation with 1000 processes is still 500 times faster than the serial SSA simulation. The full cell parallel simulation without calcium influx achieves an unprecedented 3600 times speedup with 2000 processes. This means our implementation was not only faster to finish than a serial simulation, but is also capable to compute 0.8 times more realizations of the stochastic simulation with the same amount of computing resource and time compared to a batch of serial SSA realizations.

## 4    Discussion and Future Directions

Our current parallel STEPS implementation achieves significant performance improvement and good scalability, as shown in our test results. However, as a preliminary implementation, it lacks or simplifies several functionalities that could be important to real-world simulations. These functionalities require further investigation and development in future generations of parallel STEPS.

Comparing to the serial version of STEPS, one core functionality missing from our current implementation is the ability to simulate membrane potential as well as voltage-dependent gating channels (Hepburn et al., 2013). This is the main reason why we were unable to fully simulate the stochastic spatial calcium burst model with the Purkinje sub-branch morphology in our example, but rely on the calcium influx profile extracted from previous serial simulation instead. The combined simulation of neuronal electrophysiology and molecular reaction-diffusion has recently raised interests as it bridges the gap between computational neuroscience and systems biology, and is expected to be in great use in the foreseeing future. To address such demand, we are actively collaborating with the Human Brain Project (Markram, 2012) to develop a parallel implementation of the corresponding electric field (E-Field) sub-system, which will be integrated into parallel STEPS upon its completion.

As analyzed in the results, the majority of performance speedup is contributed by the reduction of $T_{comp}$ thanks to parallel computing. Eventually $T_{idle}$ becomes the main bottleneck as it is mostly constant relative to the process count, unlike $T_{comp}$ which decreases consistently. This observation suggests two future investigation and development directions, maximizing the speedup gained from $T_{comp}$, and minimizing $T_{idle}$.

Maximizing the speedup gained from $T_{comp}$ is important to real-world research because significant performance improvement needs to be achieved with reasonable amount of computational resource. Adapting advanced algorithms and optimizing memory caching ability are two common approaches to achieve this goal. At present, we mainly focus on further optimizing memory footprint and caching ability for super-large scale simulations. In the current implementation, although reaction SSA and propensity update information are distributed, each process still stores the complete information of the biochemical model and the mesh. This noticeably affects the weak scalability of our implementation, as shown in Figure 5. The redundant information is so far required for the purpose of interfacing with other non-parallel sub-systems such as serial E-Field, but we will investigate whether the model and geometry information can be split based on the demand of individual processes.

Process load balancing plays a crucial role in determining the idle time of the simulation $T_{idle}$, and consequently the maximum speedup the simulation can achieve. In an unbalanced-loading simulation, processes will always be idle until the slowest one finishes, thus dramatically increasing $T_{idle}$. This issue is essential to stochastic spatial reaction-diffusion simulations as high concentration gradients of molecules can be observed frequently in many real-world models similar to our calcium burst model. Because molecule concentrations change significantly during simulation due to reactions and diffusion, the loading of each process may change rapidly. While adding model and initial molecule concentration information to the partitioning procedure may help to balance the loading for early simulation, the initial partitioning will eventually become inefficient as molecule concentrations change. An efficient load balancing algorithm is required to solve this problem. The solution should be able to redistribute tetrahedrons between processes automatically on the fly based on their current workloads. Efficiency is the main focus of the solution, because constantly copying tetrahedron data between processes via network communication can be extremely time consuming, and may overshadow any benefit gained from the rebalancing.

In its current status, our parallel STEPS implementation remains a great improvement on the serial SSA solution. The calcium burst simulation with Purkinje cell sub-branch morphology, dynamic calcium influx and periodic data recording is representative of the simulation condition and requirements of typical real-world research. Similar models that previously required years of simulations can now be completed within days. The shortening of the simulation cycle is greatly beneficial to research as it provides opportunities for further fine-tuning the model based on simulation results.

### Conflict of Interest

The authors declare that the research was conducted in the absence of any commercial or financial relationships that could be construed as a potential conflict of interest.

### Author Contributions

WC designed, implemented and tested the parallel STEPS described, as well as drafted the manuscript. EDS conceived of and supervised the STEPS project and helped draft the manuscript. Both authors contributed to the manuscript and read and approved the submission.

### Acknowledgments

This work was funded by the Okinawa Institute of Science and Technology Graduate University. All simulations were run on the "Sango" cluster at the Okinawa Institute of Science and Technology. We are very grateful to Iain Hepburn of the Computational Neuroscience Unit, OIST, Japan, for discussion and critical review of the initial draft of this manuscript.

### References

Andrews, S. S., and Bray, D. (2004). Stochastic simulation of chemical reactions with spatial resolution and single molecule detail. *Phys. Biol.* 1, 137–151. doi:10.1088/1478-3967/1/3/001.

Anwar, H., Hepburn, I., Nedelescu, H., Chen, W., and De Schutter, E. (2013). Stochastic calcium mechanisms cause dendritic calcium spike variability. 33, 15848–15867. doi:10.1523/JNEUROSCI.1722-13.2013.

Anwar, H., Roome, C. J., Nedelescu, H., Chen, W., Kuhn, B., and De Schutter, E. (2014). Dendritic diameters affect the spatial variability of intracellular calcium dynamics in computer models. *Front Cell Neurosci* 8, 168. doi:10.3389/fncel.2014.00168.

Balls, G. T., Baden, S. B., Kispersky, T., Bartol, T. M., and Sejnowski, T. J. (2004). A large scale monte carlo simulator for cellular microphysiology. *Proceedings of 18th International Parallel and Distributed Processing Symposium*, 42:26–30.

Coupez, T., Digonnet, H., and Ducloux, R. (2000). Parallel meshing and remeshing. *Applied Mathematical Modelling* 25, 153–175. doi:10.1016/S0307-904X(00)00045-7.

D'Agostino, D., Pasquale, G., Clematis, A., Maj, C., Mosca, E., Milanesi, L., et al. (2014). Parallel solutions for voxel-based simulations of reaction-diffusion systems. *BioMed Research International* 2014, 980501–10. doi:10.1155/2014/980501.

Dematté, L. (2012). Smoldyn on graphics processing units: massively parallel Brownian dynamics simulations. *IEEE/ACM Trans Comput Biol Bioinform* 9, 655–667. doi:10.1109/TCBB.2011.106.

Dematté, L., and Mazza, T. (2008). "On Parallel Stochastic Simulation of Diffusive Systems," in *Computational Methods in Systems Biology* Lecture Notes in Computer Science. eds.M. Heiner and A. M. Uhrmacher (Berlin, Heidelberg: Springer Berlin Heidelberg), 191–210. doi:10.1007/978-3-540-88562-7_16.

Drawert, B., Engblom, S., and Hellander, A. (2012). URDME: a modular framework for stochastic simulation of reaction-transport processes in complex geometries. *BMC Syst Biol* 6, 76. doi:10.1186/1752-0509-6-76.

Fink, S. J., Baden, S. B., and Kohn, S. R. (1998). Efficient run-time support for irregular block-structured applications. *Journal of Parallel and Distributed Computing* 50, 61–82.

Gibson, M. A., and Bruck, J. (2000). Efficient Exact Stochastic Simulation of Chemical Systems with Many Species and Many Channels. *J. Phys. Chem. A* 104, 1876–1889. doi:10.1021/jp993732q.

Gillespie, D. T. (1976). A general method for numerically simulating the stochastic time evolution of coupled chemical reactions. *Journal of Computational Physics* 22, 403–434. doi:10.1016/0021-9991(76)90041-3.

Gillespie, D. T. (2001). Approximate accelerated stochastic simulation of chemically reacting systems. *J. Chem. Phys.* 115, 1716. doi:10.1063/1.1378322.

Gladkov, D. V., Alberts, S., D'Souza, R. M., and Andrews, S. (2011). Accelerating the Smoldyn spatial stochastic biochemical reaction network simulator using GPUs. in (Society for Computer Simulation International), 151–158.

Hattne, J., Fange, D., and Elf, J. (2005). Stochastic reaction-diffusion simulation with MesoRD. *Bioinformatics* 21, 2923–2924. doi:10.1093/bioinformatics/bti431.

Hepburn, I., Cannon, R., and De Schutter, E. (2013). Efficient calculation of the quasi-static electrical potential on a tetrahedral mesh and its implementation in STEPS. *Front. Comput. Neurosci.* 7. doi:10.3389/fncom.2013.00129.

Hepburn, I., Chen, W., and De Schutter, E. (2016). Accurate reaction-diffusion operator splitting on tetrahedral meshes for parallel stochastic molecular simulations. *J. Chem. Phys.* 145, 054118–22. doi:10.1063/1.4960034.

Hepburn, I., Chen, W., Wils, S., and De Schutter, E. (2012). STEPS: efficient simulation of stochastic reaction–diffusion models in realistic morphologies. *BMC Systems Biology* 6, 36. doi:10.1186/1752-0509-6-36.

Kerr, R. A., Bartol, T. M., Kaminsky, B., Dittrich, M., Chang, J.-C. J., Baden, S. B., et al. (2008). Fast Monte Carlo Simulation Methods for Biological Reaction-Diffusion Systems in Solution and on Surfaces. *SIAM J. Sci. Comput.* 30, 3126–3149. doi:10.1137/070692017.

Koh, W., and Blackwell, K. T. (2011). An accelerated algorithm for discrete stochastic simulation of reaction–diffusion systems using gradient-based diffusion and tau-leaping. *J. Chem. Phys.* 134, 154103. doi:10.1063/1.3572335.

Lin, Z., Tropper, C., Ishlam Patoary, M. N., McDougal, R. A., Lytton, W. W., and Hines, M. L. (2015). NTW-MT. in (New York, New York, USA: ACM Press), 157–167. doi:10.1145/2769458.2769459.

Markram, H. (2012). The Human Brain Project. *Scientific American* 306, 50–55. doi:10.1038/scientificamerican0612-50.

Marquez-Lago, T. T., and Burrage, K. (2007). Binomial tau-leap spatial stochastic simulation algorithm for applications in chemical kinetics. *J. Chem. Phys.* 127, 104101–10. doi:10.1063/1.2771548.

McDougal, R. A., and Shepherd, G. M. (2015). 3D-printer visualization of neuron models. *Front. Neuroinform.* 9, 1–9. doi:10.3389/fninf.2015.00018.

Oliveira, R. F., Terrin, A., Di Benedetto, G., Cannon, R. C., Koh, W., Kim, M., et al. (2010). The Role of Type 4 Phosphodiesterases in Generating Microdomains of cAMP: Large Scale Stochastic Simulations. *PLoS ONE* 5, e11725. doi:10.1371/journal.pone.0011725.

Roberts, E., Stone, J. E., and Luthey-Schulten, Z. (2013). Lattice Microbes: high-performance stochastic simulation method for the reaction-diffusion master equation. *J. Comput. Chem.* 34, 245–255. doi:10.1002/jcc.23130.

Rodriguez, J. V., Kaandorp, J. A., Dobrzynski, M., and Blom, J. G. (2006). Spatial stochastic modelling of the phosphoenolpyruvate-dependent phosphotransferase (PTS) pathway in Escherichia coli. *Bioinformatics* 22, 1895–1901. doi:10.1093/bioinformatics/btl271.

Slepoy, A., Thompson, A. P., and Plimpton, S. J. (2008). A constant-time kinetic Monte Carlo algorithm for simulation of large biochemical reaction networks. *J. Chem. Phys.* 128, 205101. doi:10.1063/1.2919546.

Vigelius, M., Lane, A., and Meyer, B. (2011). Accelerating reaction-diffusion simulations with general-purpose graphics processing units. *Bioinformatics* 27, 288–290. doi:10.1093/bioinformatics/btq622.

Wang, B., Yao, Y., Zhao, Y., Hou, B., and Peng, S. (2009). Experimental Analysis of Optimistic Synchronization Algorithms for Parallel Simulation of Reaction-Diffusion Systems. in (IEEE), 91–100. doi:10.1109/HiBi.2009.22.

**Algorithm 1: Parallel STEPS Runtime Main Loop**

Given $p$ as the running process, and *neighs_p* as the set of processes hosting tetrahedrons that are neighbors to any tetrahedron hosted by $p$

**for** each $p'$ in *neighs_p*, **do**
    **create** remote_change_buffer$_{p'}$
**end**
**initialize** buffer_sent_complete = True

**while** $t < t_{end}$, **do**
  **if** $t + \tau > t_{end}$, **then** $\tau = t_{end} - t$

  # Reaction SSA Operator
  **initialize** $\Delta t = 0.0$
  **for** each reactant and product $S$ in each hosted reaction $R$, **do**
    **initialize** $t_S = 0.0$
    **initialize** $O_S = 0.0$
  **end**
  **while** $\Delta t \leq \tau$, **do**
    compute the next reaction time $t_{next}$ using SSA
    **if** $\Delta t + t_{reac} > \tau$ **then break**
    find the next reaction event $R_{next}$ using SSA
    **if** no $R_{next}$ is found **then break**
    $\Delta t += t_{reac}$
    apply molecule changes caused by $R_{next}$
    **for** each reactant and product species $S$ of $R_{next}$, **do**
      $O_S += N_S(\Delta t - t_S)$, where $N_S$ is the previous molecule count of $S$ at $t_S$
      $t_S = \Delta t$
    **end**

**(cont. )**

**Algorithm 1: Parallel STEPS Runtime Main Loop (conc.)**

# Diffusion Operator
**if** buffer_sent_complete == False, **then**
    wait for all buffers to be sent
    buffer_sent_complete = True
**for** each $p'$ in *neighs_p*, **do**
    **reset** remote_change_buffer$_{\rho'}$
**end**

**for** each diffusive species $S$ in tetrahedron *tet* hosted by $p$, **do**
    $\overline{N} = O_S / \tau$
    distribute $n$ molecules among neighbors of *tet*, where $n = binomial(\overline{N}, \tau, d_{S,tet})$
    **for** each neighboring tetrahedron, *neigh_tet*, **do**
        **if** *neigh_tet* is hosted by $p$ **then**      #local diffusion
            apply molecule changes for tet and *neigh_tet*
        **else if** *neibh_tet* is hosted by $p', p' \neq p$ **then**    #cross-process diffusion
            apply molecule changes for *tet*
            register molecule changes for *neigh_tet* to remote_change_buffer$_{\rho'}$
        **end if**
    **end**
**end**
update affected propensities of hosted reactions and diffusions in $p$

# Cross-Process Communication and Update
**for** each $p'$ in *neighs_p*, **do**
    send remote_change_buffer$_{\rho'}$ to $p'$ using non-blocking communication
**end**
buffer_sent_complete = False

**initialize** n_buffer_received = 0
**while** n_buffer_received $\neq$ sizeof(*neighs_p*), **do**
    wait until a buffer arrives
    receive buffer
    apply molecule changes and update propensities according to the buffer data
    n_buffer_received += 1
**end**
$t$ += $\tau$
**end**

**if** buffer_sent_complete == False, **then**
    wait for all buffers to be sent
**for** each $p'$ in *neighs_p*, **do**
    **delete** remote_change_buffer$_{\rho'}$
**end**

Table 1. Simulation Settings for Weak Scalability Study

| Geometry Dimensions | Initial Molecule Count | Num. Processes |
|---|---|---|
| $10 \times 10 \times 100 \mu m^3$ | default | 300 |
| $10 \times 10 \times 200 \mu m^3$ | 2x | 600 |
| $10 \times 20 \times 100 \mu m^3$ | 2x | 600 |
| $10 \times 10 \times 300 \mu m^3$ | 3x | 900 |
| $10 \times 30 \times 100 \mu m^3$ | 3x | 900 |

**Figures**

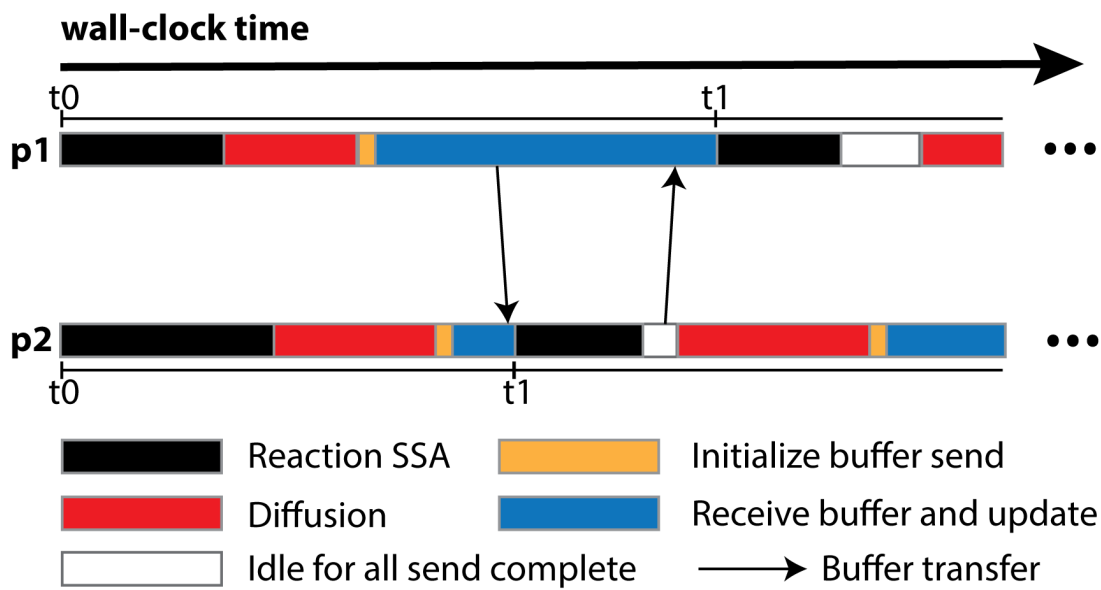

Figure 1. Schematic illustration of different runtime stages of two processes, $p_1$ and $p_2$, assuming that $p_1$ is the only neighboring process of $p_2$. Once $p_2$ receives and applies the remote change buffer from $p_1$ for iteration $t_0$, it can immediately start the reaction SSA computation for iteration $t_1$, without waiting for $p_1$ to complete iteration $t_0$.

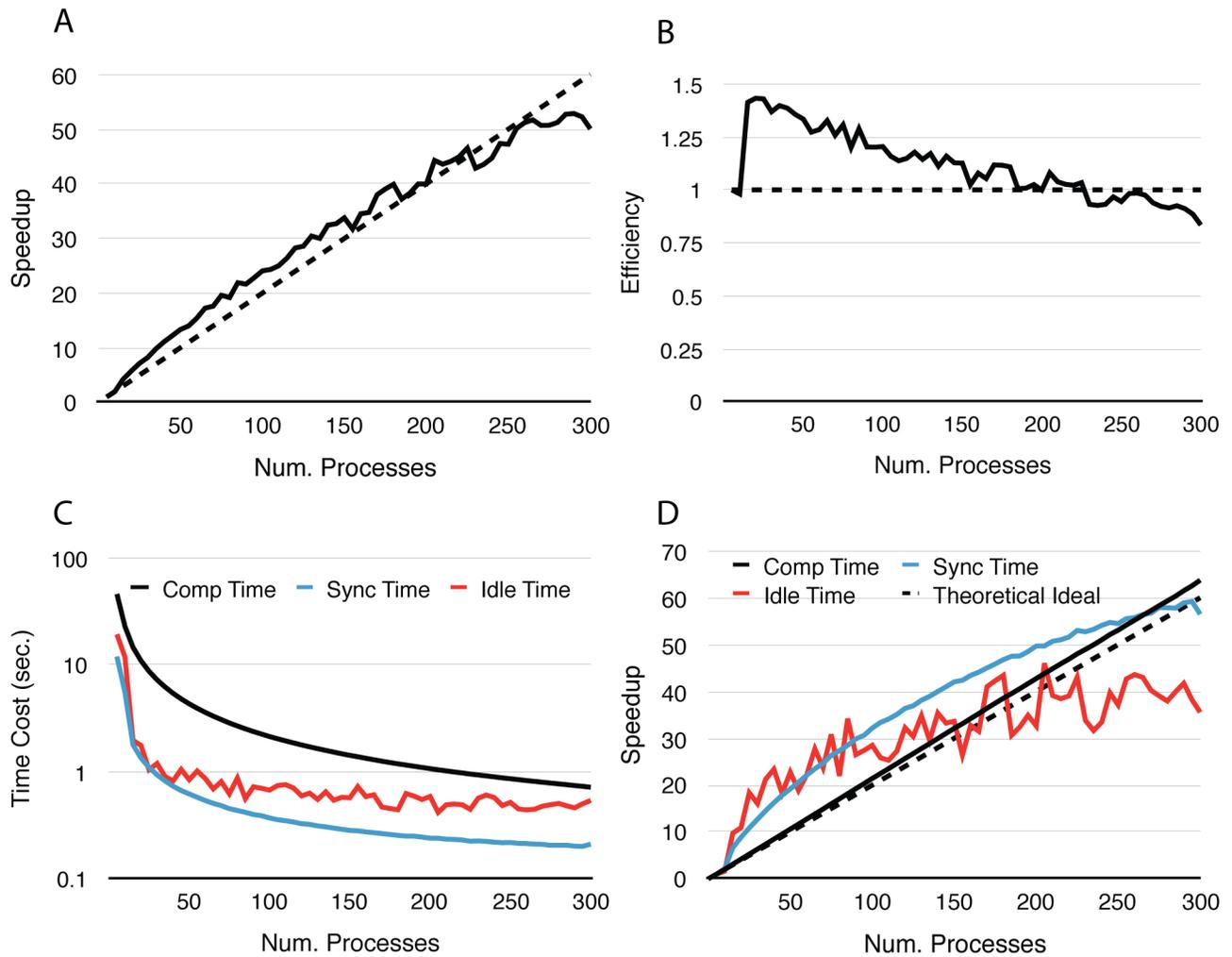

Figure 2. Performance of parallel simulations with simple model and geometry. Each series start from $p = 5$ and progressively increase to $p = 300$. Both speedup and efficiency are measured relative to simulations with $p = 5$. **(A)** Simulations maintain super-linear speedup until p ≈ 200. **(B)** In general efficiency decreases as $p$ increases, but remains above 0.8 in the worst case ($p = 300$). **(C** and **D)** $T_{Comp}$ contributed the majority of the speedup as it is the most time consuming segment during simulation, it maintains super-linear speedup throughout the whole series. However, as $T_{Comp}$ decreases, $T_{idle}$ becomes a critical factor because its change is insignificant once $p$ is above 100.

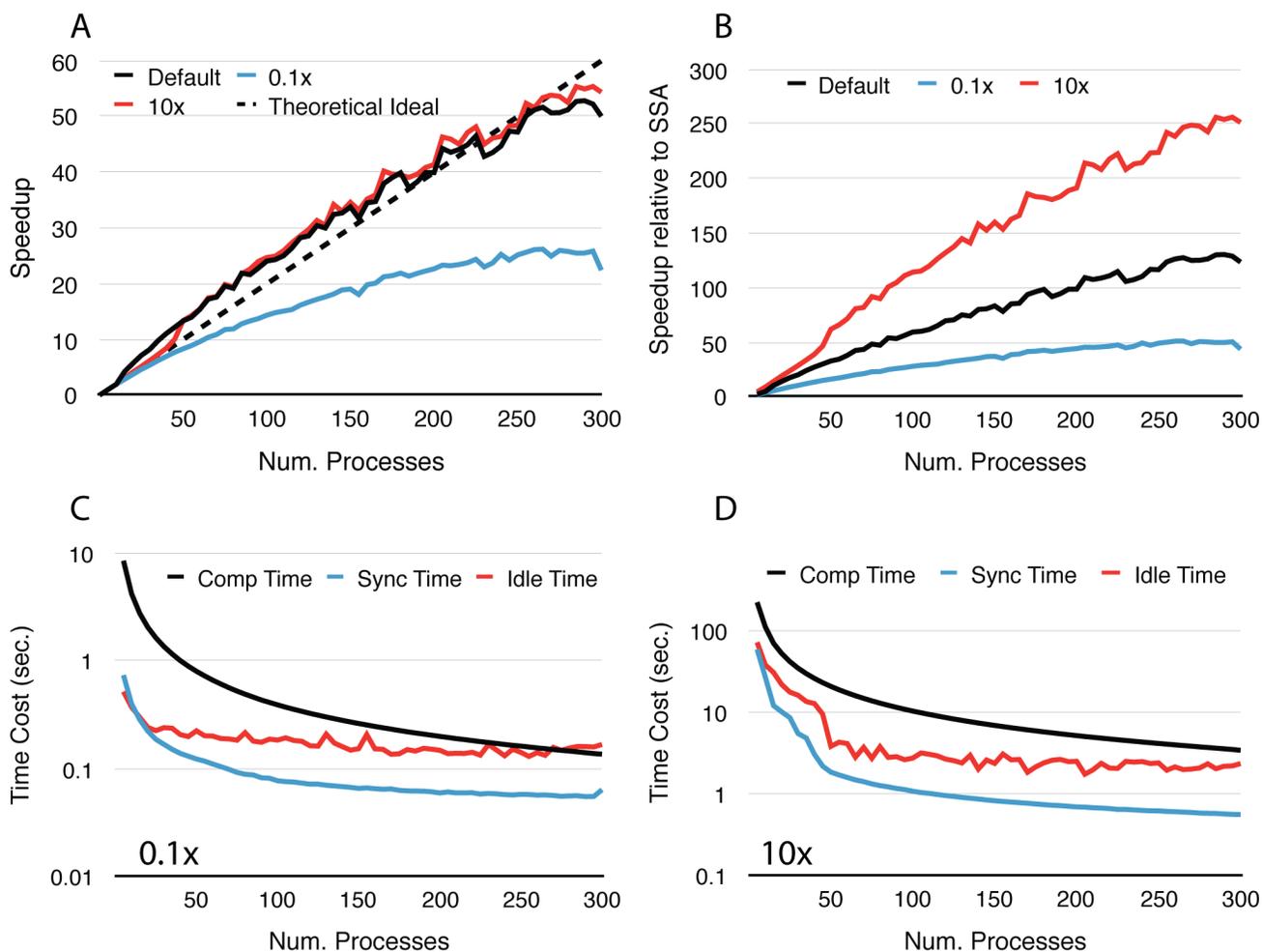

Figure 3. Performances of simulations with different molecule density. **(A)** Speedups relative to simulations with $p = 5$. Simulations with low molecule density (0.1x) achieve smaller speedups compared to the default and high density (10x) cases. **(B)** In general, simulation with higher molecule density and larger scale of parallelization achieves higher speedup relative to the serial SSA counterpart. **(C)** In the 0.1x cases, $T_{comp}$ rapidly decreases and eventually drops below $T_{idle}$, thus the overall speedup is less significant. **(D)** In the 10x cases, $T_{comp}$ remains above $T_{idle}$, therefore its contribution to speedup is significant throughout the series.

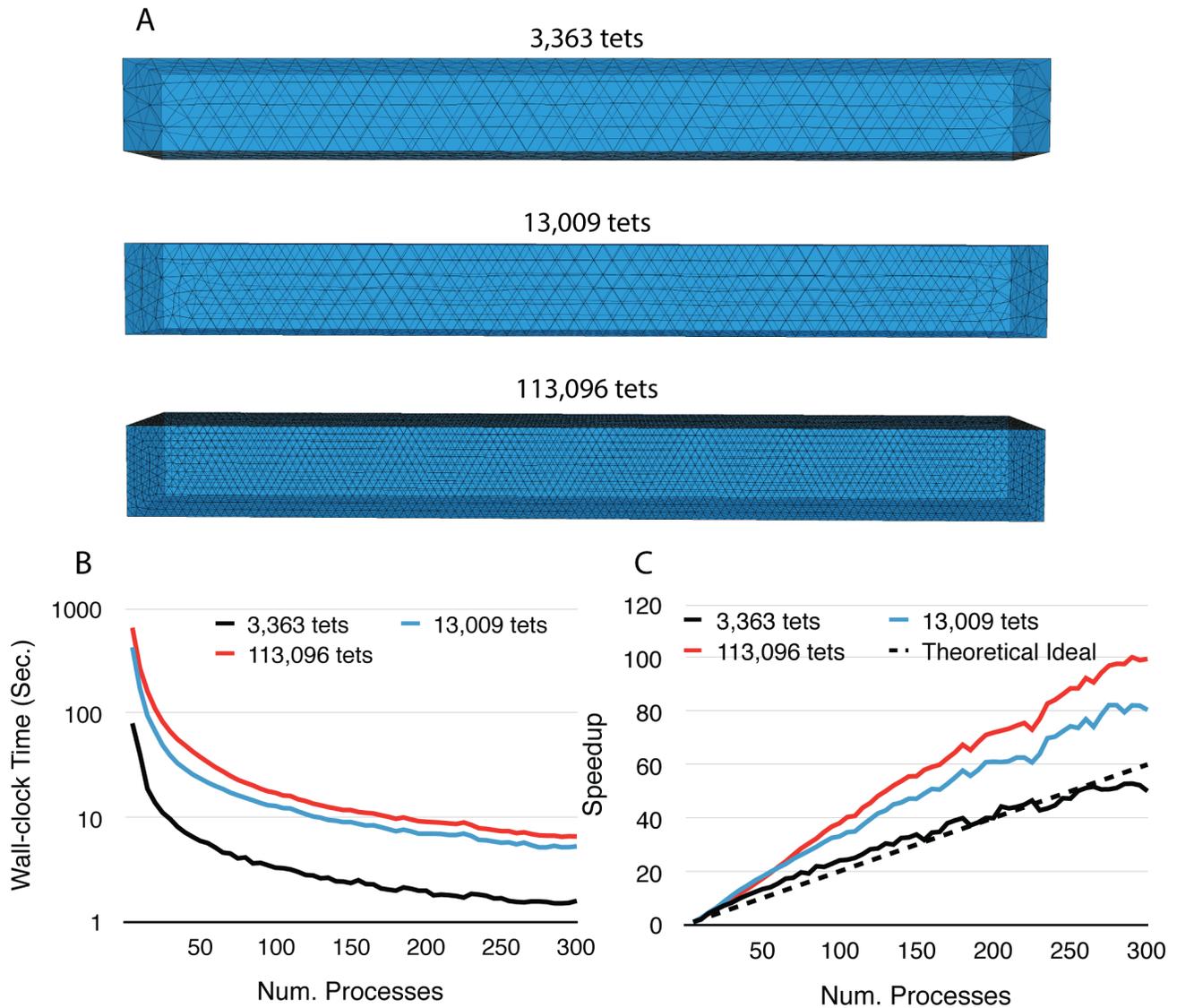

Figure 4. Performances of simulations with different mesh coarseness. **(A)** Meshes with the same geometry dimensions but different number of tetrahedrons are simulated. **(B)** Simulation with finer mesh takes longer to complete, but the time increment is not directly proportional to the increment of tetrahedrons. **(C)** Speedups relative to $T_5$. Simulation with finer mesh achieves much higher speedup in massive parallelization, thanks to the memory caching effect.

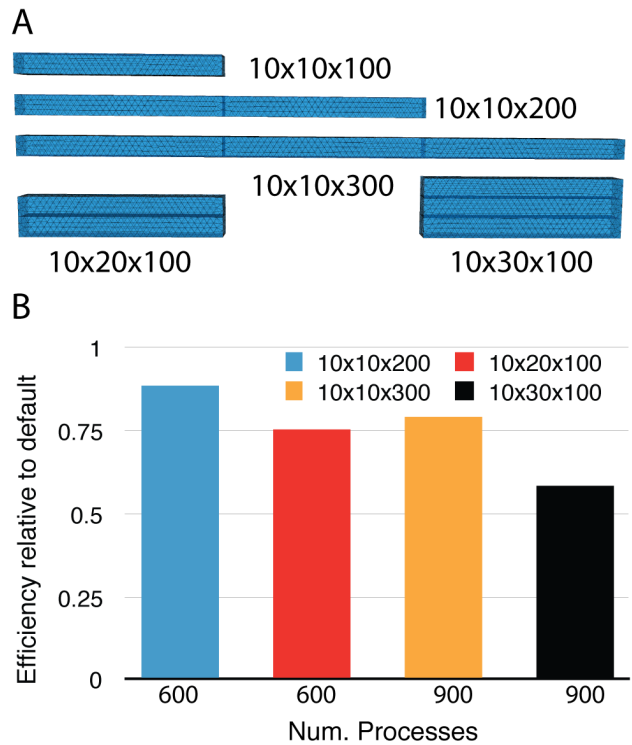

Figure 5. Weak scalability of the implementation. **(A)** The default $10{\times}10{\times}100\mu m^3$ mesh is expended along either the y or z axis as problem size increases. **(B)** Efficiencies relative to the default case ($p = 300$).

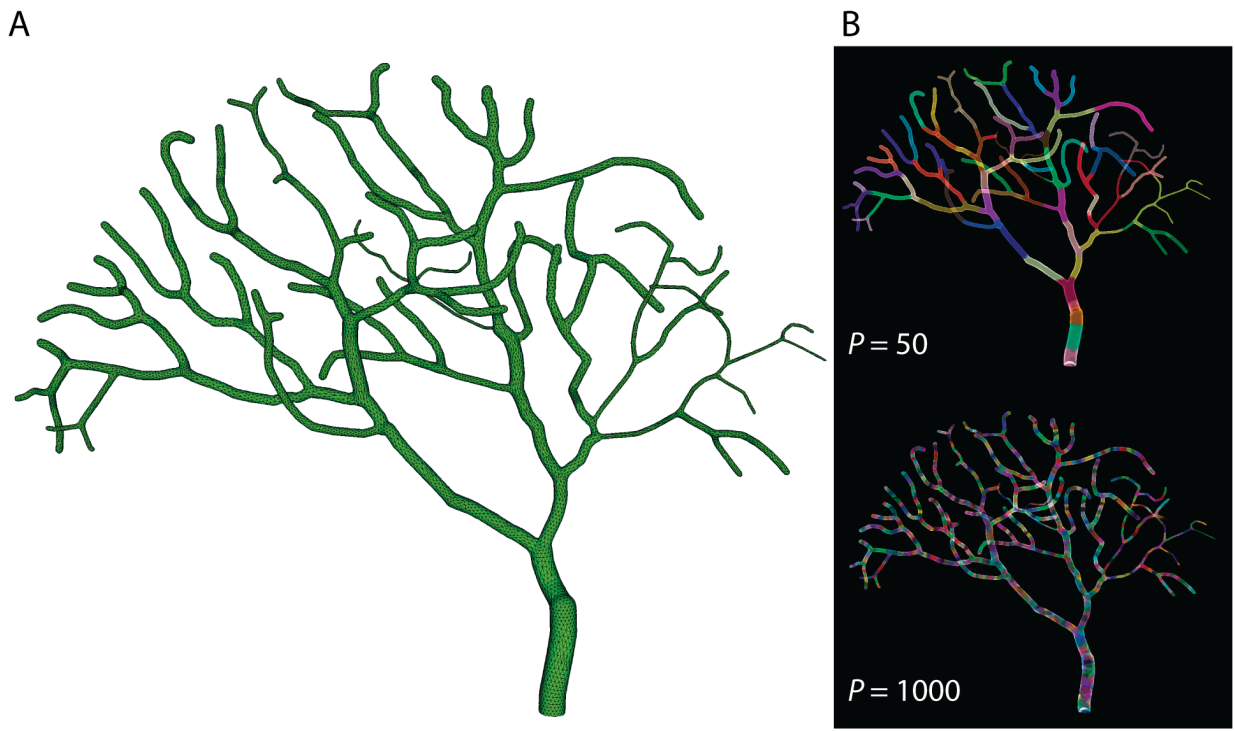

Figure 6. **(A)** Tetrahedral mesh of a Purkinje cell sub-branch morphology. This mesh consists of 111,664 tetrahedrons. **(B)** Partitioning generated by Metis for $p = 50$ and $p = 1000$. Each color segment indicates a set of tetrahedrons hosted by a single process.

A

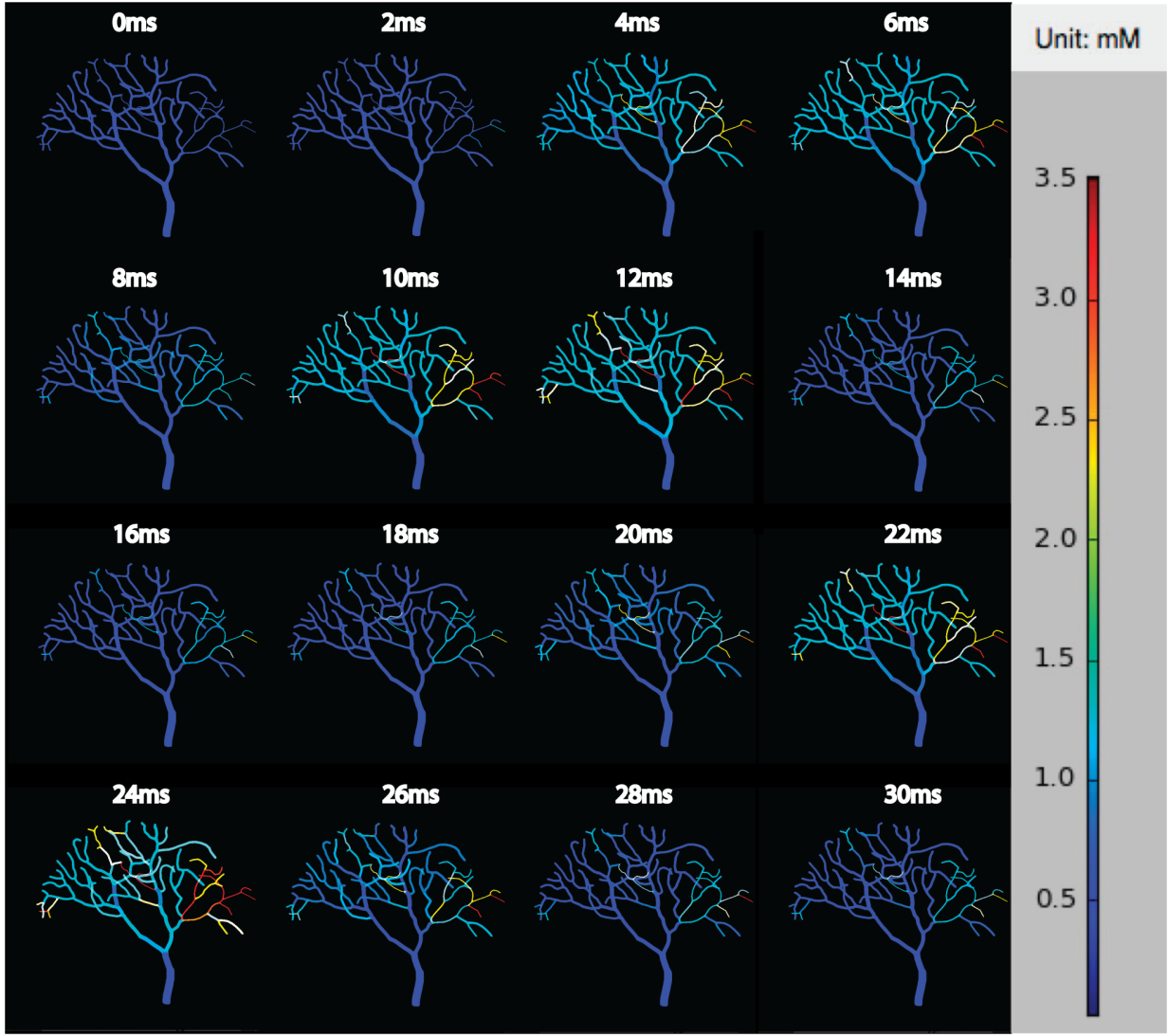

B

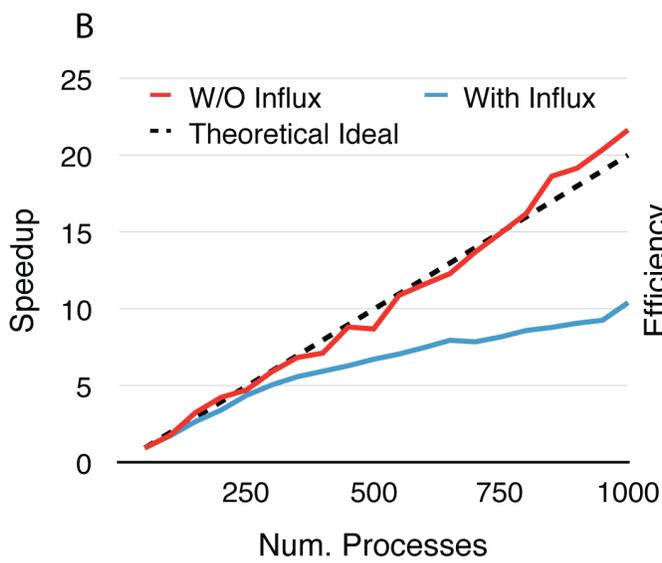

C

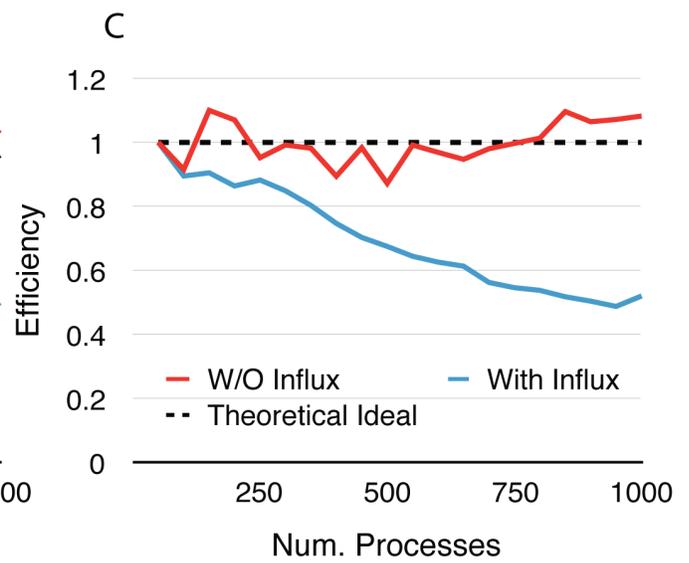

Figure 7. Calcium burst simulations with a Purkinje cell sub-branch morphology. **(A)** Calcium activity of each branch segment over a single trail period, visualized by the STEPS visualization toolkit. The calcium activity shows large spatial and temporal divergences, which significantly affects the speedup **(B)** and efficiency **(C)** of the simulation.

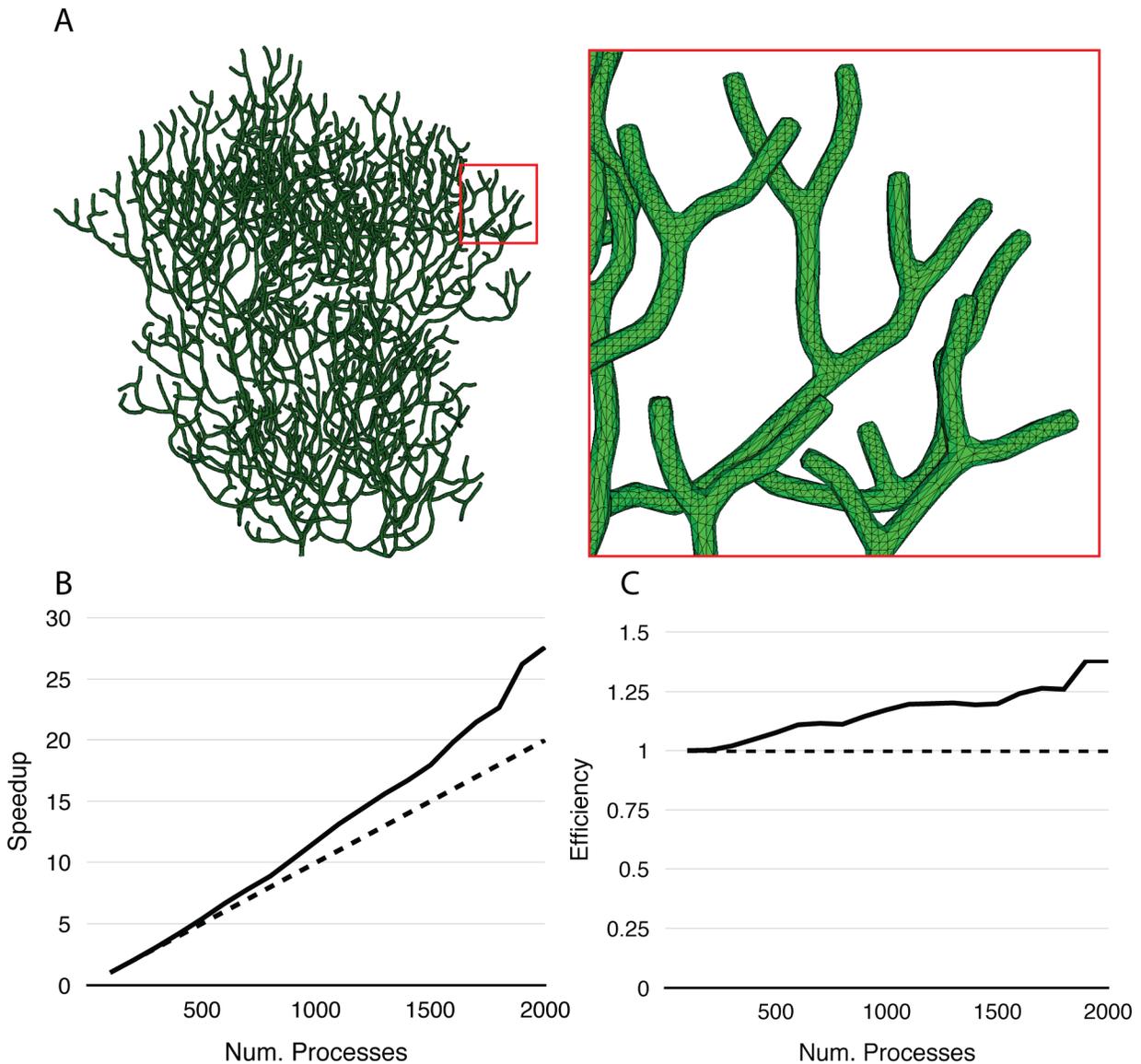

Figure 8. Performance of reaction-diffusion simulation with a mesh of a complete Purkinje dendrite tree. **(A)** Morphology of the mesh and a zoom-in look of its branches. The mesh consists of 1,044,155 tetrahedrons. **(B)** Speedup relative to the simulation with $p = 100$ shows super-linear scalability. **(C)** Efficiency also increases as $p$ increases, indicates that better efficiency may be achieved with more processes.

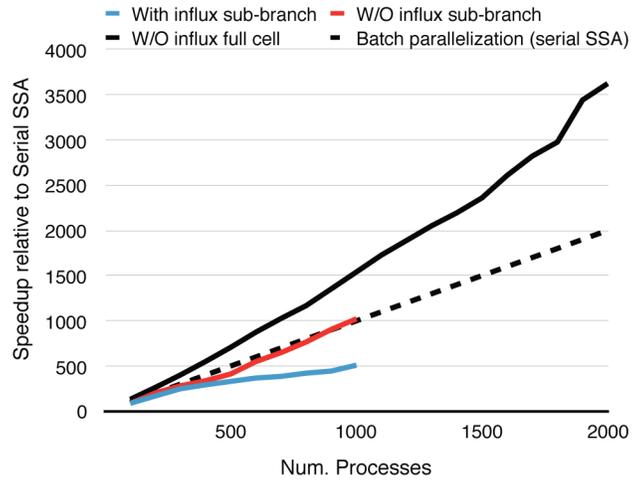

Figure 9. Speedups of the parallel calcium burst simulations relative to their serial SSA counterparts. Dashed line assumes that *p* processes are used to simulate a batch of *p* serial SSA realizations.